\documentclass[twocolumn,aps,showpacs,prb,tightenlines,amsmath,amssymb]{revtex4}
\usepackage{graphicx}
\usepackage{amssymb}
\usepackage{amsmath}
\usepackage{colordvi}
\begin{document}

\title{Non-Markovian hole spin kinetics in $p$-type GaAs quantum wells}
\author{P. Zhang}
\affiliation{Hefei National Laboratory for Physical Sciences at
  Microscale, University of Science and Technology of China, Hefei,
  Anhui, 230026, China}
\affiliation{Department of Physics,
  University of Science and Technology of China, Hefei,
  Anhui, 230026, China}
\author{M. W. Wu}
\thanks{Author to whom correspondence should be addressed}
\email{mwwu@ustc.edu.cn.}
\affiliation{Hefei National Laboratory for Physical Sciences at
  Microscale, University of Science and Technology of China, Hefei,
  Anhui, 230026, China}
\affiliation{Department of Physics,
  University of Science and Technology of China, Hefei,
  Anhui, 230026, China}
\altaffiliation{Mailing Address}

\date{\today}
\begin{abstract}

Based on fully microscopic
kinetic spin Bloch equation approach, we show the non-Markovian effect
in spin dephasing of heavy holes in $p$-type GaAs quantum wells.
The non-Markovian effect manifests itself in spin
dephasing when the mean spin precession time
is shorter than the momentum relaxation time.
The spin dephasing becomes slower  when the non-Markovian effect is
important. Moreover, quantum spin beats due to the memory effect
of the non-Markovian hole-longitudinal optical phonon scattering
are predicted.
\end{abstract}

\maketitle

Much attention has been devoted to the spin degree of freedom of
carriers in  semiconductors recently for the purpose of the possible
application of spintronic devices.\cite{wolf,ziese,aws,jaro}
Understanding spin relaxation/dephasing  is one of the most
important issues.\cite{meier} Recently an extensive investigation
has been performed based on kinetic spin Bloch equations
(KSBEs)\cite{wu-rev} to understand the spin relaxation/dephasing in
zinc-blende semiconductors.\cite{wu1,wu2,wu3,weng,lv} In these
works, together with earlier studies,\cite{meier} the kinetics was
treated in Markovian limit. The Markovian limit is a good
approximation in the strong scattering limit where the spin
precession time ${\Omega}^{-1}$ is long compared to the momentum
relaxation time $\tau_{p}$, i.e., ${\Omega}^{-1}\gg\tau_{p}$,
which is mostly the case for $n$-type GaAs quantum wells
(QWs)\cite{wu3} unless at very low temperature.\cite{Harley}
However, for hole systems (especially heavy hole systems)
 the situation changes a lot due to the strong
spin-orbit coupling strength. Here the mean spin precession
time becomes comparable with the
momentum relaxation time, i.e., ${\Omega}^{-1}\lesssim\tau_{p}$. 
Then Markovian approximation is not appropriate
any more. 
In this work we are going to extend our previous
kinetic spin Bloch equation approach\cite{wu-rev} to the non-Markovian
limit to investigate the spin dephasing of
heavy holes in $p$-type GaAs QWs. In non-Markovian kinetics, the
energy conservation of the scattering is lifted and there is
 memory effect due to the history dependence
of scattering. In fact, the non-Markovian effect has been extensively explored
in the ultrashort-pulse  spectroscopy in semiconductor
 optics.\cite{tran,Banyai} Recently it has been explored in
quantum dot system for spin dephasing induced by hyperfine
interaction.\cite{coish} Glazov and Sherman also investigated
the electron spin relaxation in QWs with strong magnetic
field by Monte Carlo simulation.\cite{glazov}

We start our investigation from a $p$-type GaAs (100) QW with well
width $a$ being small enough so that the heavy-hole and the light-hole
bands can be treated separately.\cite{lv,pk,dvb} We only consider the
heavy hole in this report. Moreover, only the lowest subband is considered.
The KSBEs constructed by the nonequilibrium Green
function method read\cite{wu1,wu2,wu3}
\begin{equation}
  \dot{\rho}_{{\bf k}}=\dot{\rho}_{{\bf k}}|_{coh}
+\dot{\rho}_{{\bf k}}|_{scat}\ ,
\end{equation}
in which $\rho_{{\bf k}}$ represent the
density matrices  of heavy hole with
 momentum {\bf k}. The diagonal terms
$\rho_{{\bf k},\sigma\sigma}\equiv f_{{\bf
    k}\sigma}$ ($\sigma=\pm3/2$) represent the hole
distribution functions and the off-diagonal ones $\rho_{{\bf
    k},\frac{3}{2}-\frac{3}{2}}
=\rho_{{\bf k},-\frac{3}{2}\frac{3}{2}}^{\ast}$
 describe the inter-spin-band
correlations for the spin coherence. The coherent terms
$\dot{\rho}_{\bf k}|_{coh}$ describe the coherent spin precessions
around the effective magnetic field from the Rashba terms ${\bf
  \Omega(k)}$ and its expressions can be found
 in Ref.\ \onlinecite{lv}. In the present report, we only
consider the heavy hole-longitudinal optical (LO) phonon
scattering in $\dot{\rho}_{\bf k}|_{scat}$ which is written in the
non-Markovian limit
\begin{eqnarray}
  \left.\frac{\partial\rho_{\bf k}}{\partial
      t}\right|_{scat}&=&\frac{1}{\hbar}\sum_{{\bf k^{\prime}}q_{z}}
g_{{\bf k-k^{\prime}},q_{z}}^{2}[S_{\bf kk^{\prime}}(t)
-S_{\bf k^{\prime}k}(t)\nonumber\\
&&\mbox{} +S_{\bf kk^{\prime}}^{\dagger}(t)-S_{\bf
    k^{\prime}k}^{\dagger}(t)]\ ,
\label{eq1}
\end{eqnarray}
with
\begin{eqnarray}
\nonumber
S_{\bf kk^{\prime}}(t)&=&e^{i\left(\omega_{0}
-\frac{E_{{\bf k}^{\prime}}-E_{\bf
    k}}{\hbar}\right)t}\int_{-\infty}^{t}d\tau [N^{>}\rho_{\bf
    k^{\prime}}^{<}(\tau)\rho_{\bf k}^{>}(\tau)
       \\ &&-N^{<}\rho_{\bf k^{\prime}}^{>}(\tau)\rho_{\bf
    k}^{<}(\tau)]e^{-i\left(\omega_{0}
-\frac{E_{\bf k^{\prime}}-E_{\bf k}}{\hbar}\right)\tau}\ .
\label{eq2}
\end{eqnarray}
Here $\rho_{\bf k}^{<}=\rho_{\bf k}$ and $\rho_{\bf k}^{>}=1-\rho_{\bf k}$.
$N^{<}=N_{0}$ and  $N^{>}=N_{0}+1$ with
$N_{0}=\left(e^{\hbar\omega_{0}/k_{B}T}-1\right)^{-1}$
standing for the Bose distribution of the LO phonons
  with frequency $\omega_{0}$ at temperature $T$. $E_{\bf
    k}=\hbar^{2}{\bf k}^{2}/2m^\ast$ is the energy of hole with wave vector
  ${\bf k}$ and effective mass $m^\ast$. $g_{{\bf
    k-k^{\prime}},q_{z}}$ is the hole-phonon interaction matrix
element.\cite{lv}

It is noted that the KSBEs in non-Markovian limit are
integro-differential equations.
The integral terms in $\left.\frac{\partial\rho_{\bf k}}{\partial
    t}\right|_{scat}$  indicate
the history dependence of scattering processes which lead to the memory
effect, with respect to that all the density matrices
enter only at the earlier time
  $\tau$. However, if $\rho_{\bf k}$ change slowly on the time scale of
collision, one can disregard the retardation and pull
 the distributions out of the time integral. In this
  completed-collision approximation, one can simply get in the
  scattering term $\delta$-functions denoting
  energy-conservation. In this way one comes to the Markovian limit, with
  memory effect lifted.
 For comparison, the KSBEs will be solved in both Markovian
and non-Markovian limits.

The integro-differential KSBEs in non-Markovian limit can be transformed into
a larger set of differential equations by factorizing the
integral terms and making time derivative actions on them.\cite{haug}
 Finally the group of equations to be solved
numerically are written into
\begin{eqnarray}
\nonumber
\frac{\partial \rho_{\bf k}}{\partial t}&=&\left.\frac{\partial \rho_{\bf k}}{\partial t}\right|_{coh}+\frac{1}{\hbar}
\sum_{{\bf k^{\prime}}q_{z}}g_{{\bf k-k^{\prime}},q_{z}}^{2}[S_{\bf kk^{\prime}}(t)-S_{\bf k^{\prime}k}(t) \\ && +S_{\bf kk^{\prime}}^{\dagger}(t)-S_{\bf
    k^{\prime}k}^{\dagger}(t)]\ ,\\ \nonumber
\frac{\partial S_{\bf kk^{\prime}}}{\partial t}&=&
i[\omega_{0}-(E_{\bf
        k^{\prime}}-E_{\bf k})/\hbar] S_{\bf
      kk^{\prime}}\\ &&+[N^{>}\rho_{\bf
    k^{\prime}}^{<}(t)\rho_{\bf k}^{>}(t)-N^{<}\rho_{\bf
    k^{\prime}}^{>}(t)\rho_{\bf k}^{<}(t)]\ .
\label{ep3}
\end{eqnarray}
Once the KSBEs are solved numerically, the temporal evolution of the
hole distribution $f_{{\bf k}\pm\frac{3}{2}}(t)$
and the spin coherence $\rho_{{\bf k}\frac{3}{2}-\frac{3}{2}}(t)$
are obtained. As discussed in the
previous papers,\cite{wu1,wu2,haug,kuhn} the spin dephasing can be obtained
from the slope of the envelop of the incoherently summed spin coherence
$\rho_{HH}(t)=\sum_{\bf  k}|\rho_{{\bf k}\frac{3}{2}-\frac{3}{2}}(t)|$.

\begin{figure}[ht]
 \begin{minipage}[]{10cm}
    \hspace{-2.5cm}\parbox[t]{10cm}{
      \includegraphics[width=8cm]{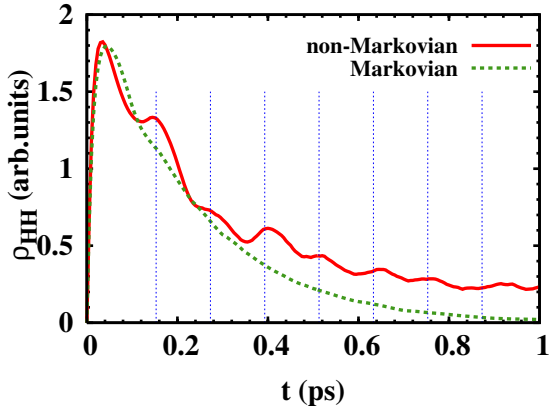}}
   \end{minipage}
  \caption{(Color online) Time evolution of the incoherently
summed spin coherence  $\rho_{HH}$. Solid curve: in non-Markovian
limit; Dashed curve: in Markovian limit.}
  \label{fig1}
\end{figure}

\begin{figure}[ht]
  \begin{minipage}[]{10cm}
    \hspace{-1.5cm}\parbox[t]{5cm}{
      \includegraphics[width=4.5cm]{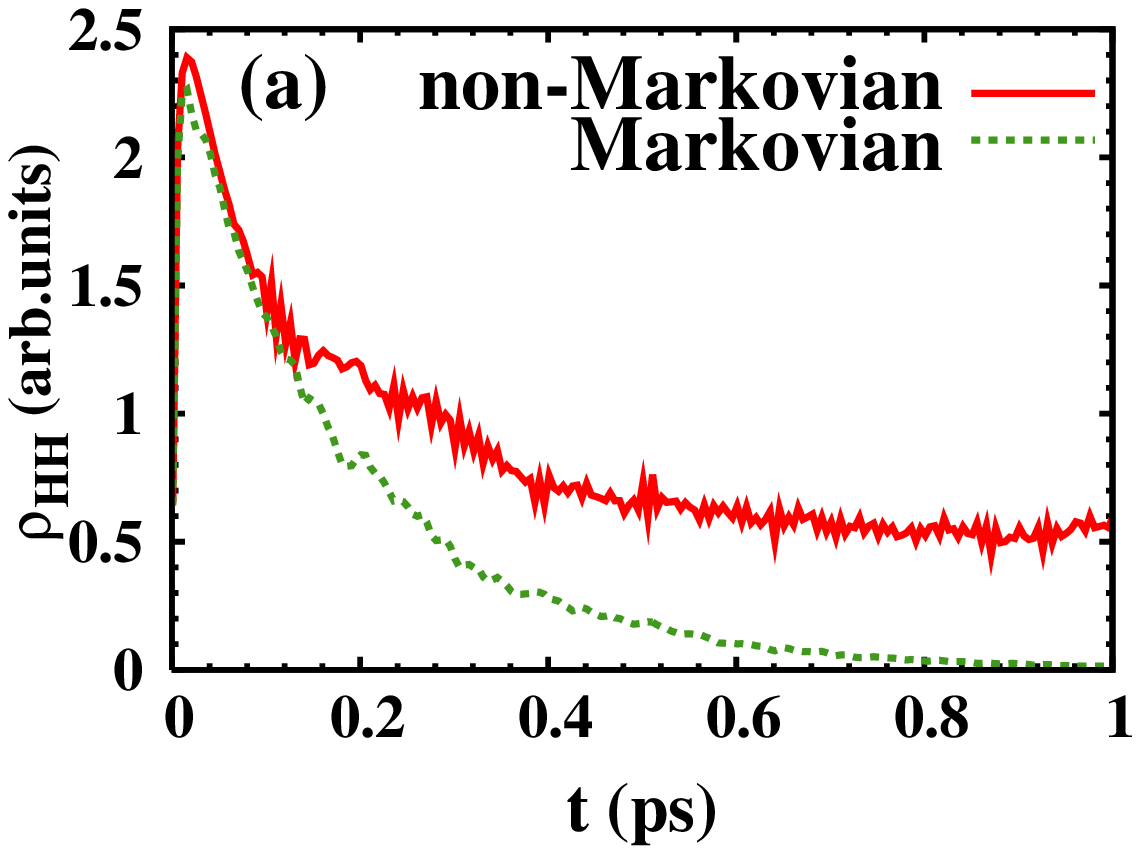}}
    \hspace{-0.5cm}\parbox[t]{5cm}{
      \includegraphics[width=4.5cm]{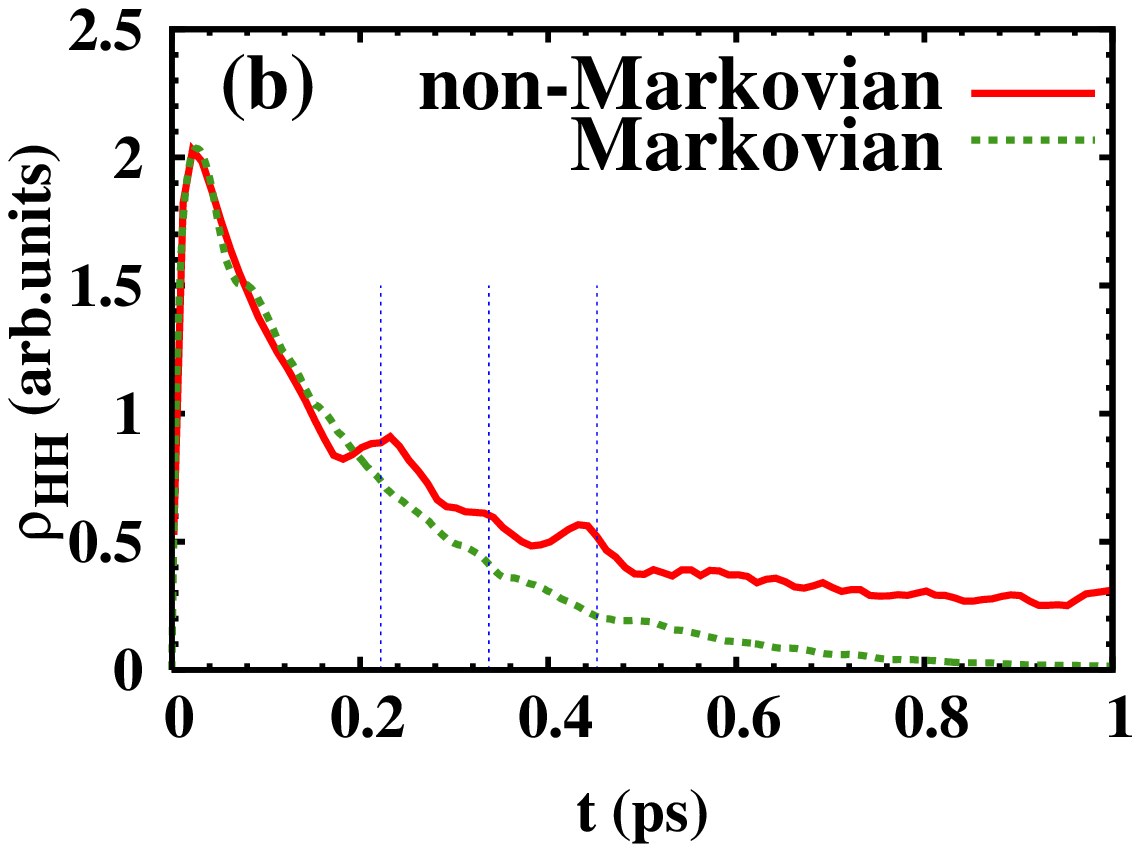}}
  \end{minipage}
  \begin{minipage}[]{10cm}
    \hspace{-1.5cm}\parbox[t]{5cm}{
      \includegraphics[width=4.5cm]{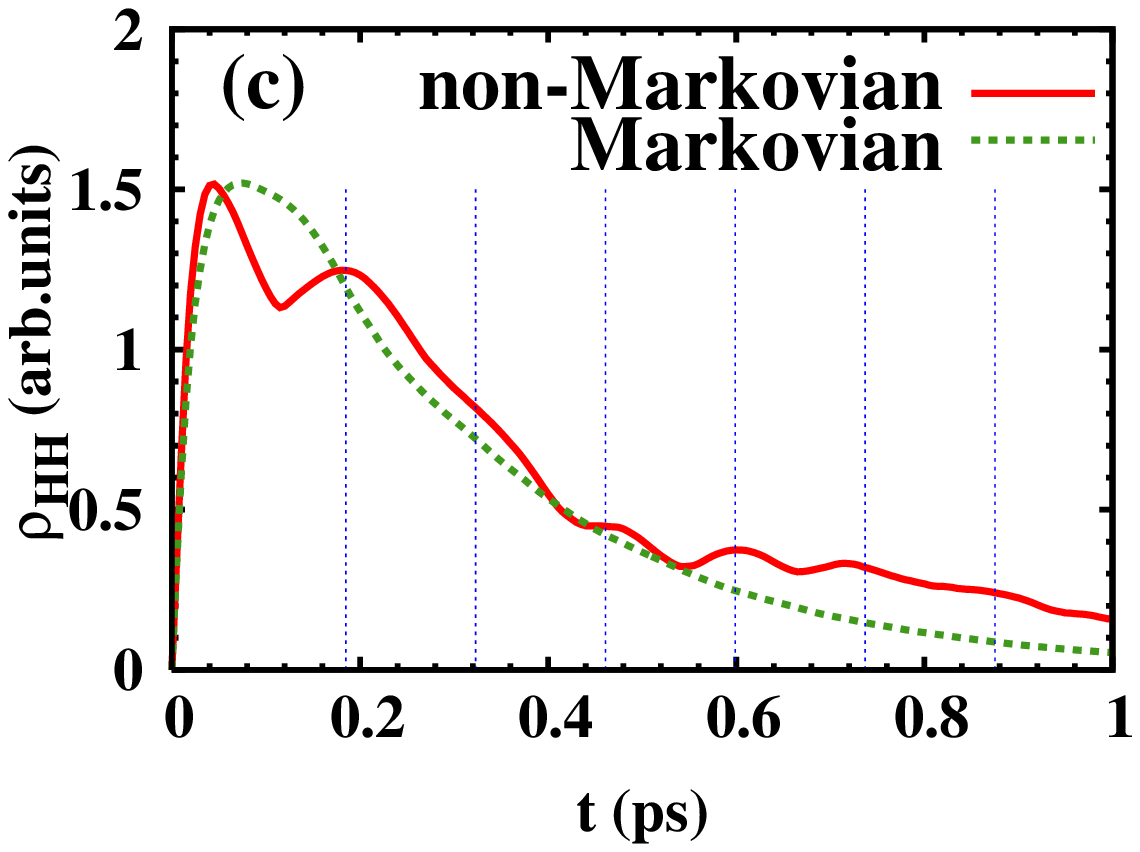}}
    \hspace{-0.5cm}\parbox[t]{5cm}{
      \includegraphics[width=4.5cm]{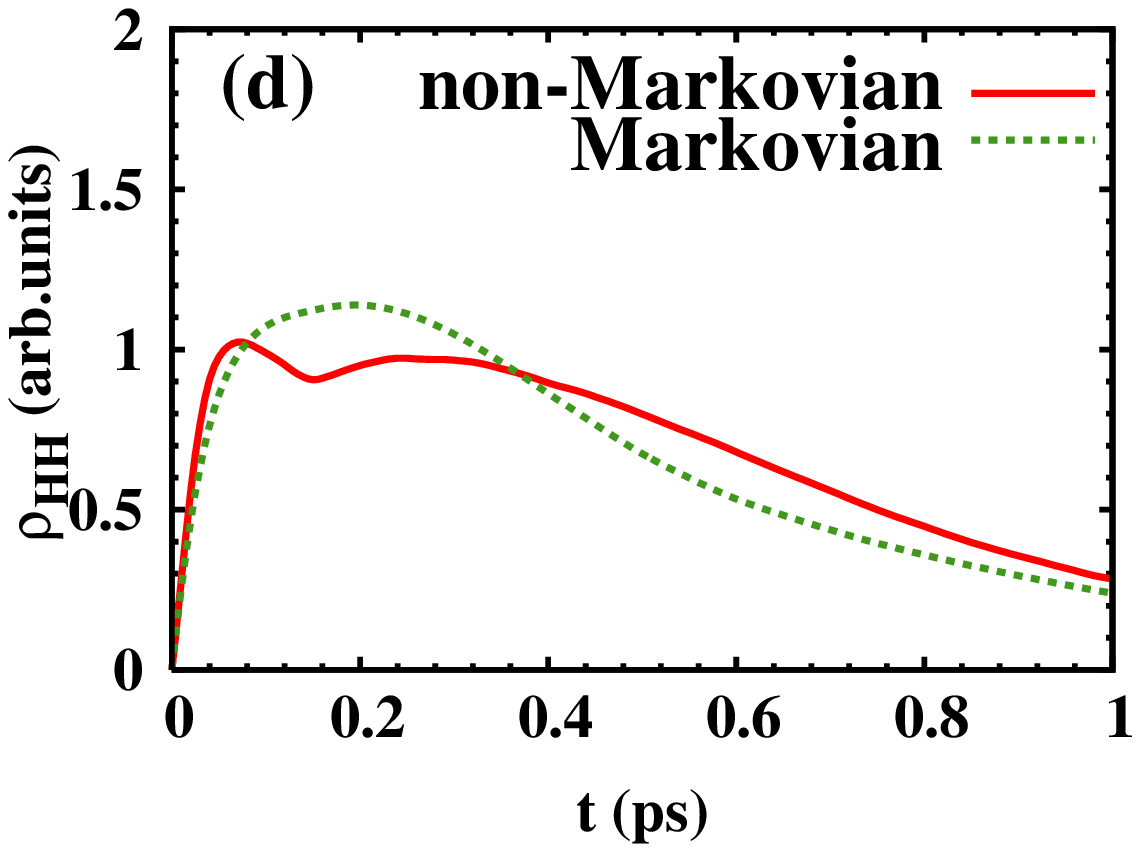}}
  \end{minipage}
  \caption{(Color online)  Time evolutions of the incoherently
summed spin coherence  $\rho_{HH}$ at scaling
parameters $\chi=5$ (a); 2 (b); 0.5 (c) and
0.2 (d), respectively. Correspondingly, the mean spin
precession time ${\Omega}^{-1}$ is 0.046\ ps (a); 0.116\ ps
(b); 0.463\ ps (c) and 1.160\ ps (d). Solid curve: in non-Markovian
limit; Dashed curve: in Markovian limit.
}
  \label{fig2}
\end{figure}

We solve the KSBEs numerically and the main results are plotted in
Figs.\ \ref{fig1} and \ref{fig2}. In the calculation,
 $T=300$\ K and  $a=5$\ nm. The hole
density $N=N_{+}+N_{-}=4\times10^{15}$\ m$^{-2}$ and the initial spin
polarization $P=(N_{+}-N_{-})/(N_{+}+N_{-})=0.025$ with $N_{+}$
($N_{-}$) standing for the density of holes with up- (down-) spin.
The material parameters of GaAs QWs can be found in
Ref.\ \onlinecite{lv} except that the
heavy hole effective mass $m^\ast$, deduced from the 8$\times$8 Luttinger-Kohn
Hamiltonian based on the quasi-degenerate theory,\cite{winkler}
depends on the width of the QW and is
 0.237$m_{0}$ when $a=5$\ nm.
We select such a value of $E_{z}$ that it satisfies
$\gamma_{54}^{7h7h}E_{z}m_{0}/\hbar^2=1.25$\ nm in the Rashba term when
$a=5$\ nm.\cite{lv}

In Fig.\ \ref{fig1} the time evolutions of the
 incoherently summed spin coherence in both the Markovian and
non-Markovian limits are plotted. One can see that the spin dephasing
is slower in the
 non-Markovian limit than the  Markovian one. This is in agreement with
the result of Glazov and Sherman\cite{glazov} and can be interpreted as the
influence of the memory effect---in non-Markovian kinetics,
the spin coherence of heavy
holes can be partially kept due to uncompleted hole-LO phonon
scattering processes. It is interesting to note that
there are quantum spin beats superimposed in the
decaying signal in the non-Markovian spin kinetics with the beat
period being the period of the LO phonon ($2\pi\omega_0^{-1}\approx0.115$\ ps).
This fascinating phenomenon is deemed as the characteristic feature of the
non-Markovian effect, similar to the
quantum beats in the ultrafast four-wave mixing signals in
GaAs.\cite{tran,Banyai} The beating is caused by the transfer of
 spin coherence of holes between  uncompleted
scattering---when a phonon is partially emitted, it can be absorbed
back by the hole system without losing  coherence. One can
realize that this can not happen in
the Markovian kinetics, where all the scattering processes are instantaneous
and complete without feeding back.

In order to gain more insight into the condition for  non-Markovian
effect, we repeat the calculations with different mean spin precession
times. To do this, we introduce a scaling parameter $\chi$ in the
Rashba term, i.e., $\chi{\bf \Omega}_{\bf k}$.
Experimentally the value of the Rashba coefficient can be tuned by bias
voltage.\cite{winkler1,papadakis}
The results are plotted in Fig.\ \ref{fig2}.
It is shown from the figure that
the smaller the mean precession time
$\Omega^{-1}\equiv 2\pi\hbar\langle{\chi
  \Omega}_{\bf k}\rangle^{-1}$
is, the more pronounced the
non-Markovian effect becomes.
Here\cite{lv}
\begin{equation}
{\langle{\Omega_{\bf k}}\rangle}
\equiv\frac{\int_{0}^{+\infty}dE_{\bf k}(f_{{\bf
  k}\frac{3}{2}}-f_{{\bf k}-\frac{3}{2}}){\Omega_{\bf k}}}{
\int_{0}^{+\infty}dE_{k}(f_{{\bf  k}\frac{3}{2}}-f_{{\bf
    k}-\frac{3}{2}})}\ .
\end{equation}
When $\chi=0.2$ for which $\Omega^{-1}$ is much
longer than the momentum relaxation time $\tau_{p}$,\cite{lei}
the non-Markovian curve approaches the
Markovian one ($\Omega^{-1}$ is about 0.232\ ps when
$\chi=1$ and $\tau_{p}$ is about 0.121\ ps).
This suggests that when
$\Omega^{-1}\gg\tau_{p}$, the Markovian approximation is a good approximation, as
expected. However, when $\Omega^{-1}$ decreases, the
non-Markovian curves deviate from the Markovian ones and become
more pronounced for smaller $\Omega^{-1}$,
indicating stronger non-Markovian effect.
Moreover, it is noted from the figure that strong  quantum spin beats
appear when the mean spin
precession time is comparable with the momentum relaxation time.
When the mean spin precession time is too short, i.e., the
inhomogeneous broadening\cite{wu-rev} is too strong, the
 quantum spin beats are
smeared out as shown in Fig.\ \ref{fig2} (a) when $\chi=5$.

In conclusion, we investigate the non-Markovian spin kinetics of
  heavy holes in $p$-type GaAs QWs.
The result is compared
with the Markovian spin kinetics. We show that the
non-Markovian effect---slower spin dephasing appears when the
inhomogeneous broadening is strong (i.e., the mean
spin precession time is shorter than the momentum
relaxation time). We further predict quantum spin beats due to the
memory effect of the non-Markovian hole-LO phonon scattering in the
spin signals. Strong quantum beats appear when the mean
spin precession time is comparable to the momentum
relaxation time.

 This work was supported by the Natural Science Foundation of China
  under Grant Nos.\ 10574120 and 10725417,  the National Basic Research Program of
China under Grant No.\ 2006CB922005,
the Innovation Project of Chinese Academy of Sciences and SRFDP.
One of the authors (P.Z.) would like to thank
J. H. Jiang, C. L\"u, and J. L. Cheng for valuable discussions.

\end{document}